\documentclass[a4paper,12pt]{article}
\usepackage{amstex,righttag,epsfig,rotating}
\begin{document}
\begin{titlepage}
\begin{flushright}
Z\"urich University Preprint\\
ZU-TH 33/96\\
gr-qc/9703052\\
\end{flushright}
\begin{center}
\vfill
{\large\bf Time evolution of the perturbations for a complex scalar field in a 
Fried\-mann\--Le\-ma\^{\i}\-tre universe$^*$}
\vfill
{\bf Philippe Jetzer$^{1,2}$ and David Scialom$^{2}$}
\vskip 0.5cm
$^1$ Paul Scherrer Institute, Laboratory for Astrophysics, CH-5232 Villigen
PSI \\
$^2$ Institute of Theoretical Physics, University of Z\"{u}rich, 
\\Winterthurerstrasse
190, CH-8057 Z\"{u}rich, Switzerland
\end{center}
\vfill
\begin{abstract}
We study the time evolution of small classical perturbations
in a gauge    
invariant way for a complex
scalar field in the early zero curvature Friedmann-Lema\^{\i}tre
universe. We, thus, generalize the analysis which has been done so
far for a real scalar field. 
We give also a derivation of the Jeans wavenumber
in the Newtonian regime starting from the general
relativistic equations, avoiding the so-called Jeans swindle.

During the inflationary phase, whose length depends on the value of
the bosonic charge, the behavior of the
perturbations turns out to be the same as for a real
scalar field. In the
oscillatory phase the time evolution of the
perturbations can be determined analytically as long as the bosonic
charge of the corresponding
background solution is sufficiently large. This is not
possible for the real scalar field, since the corresponding bosonic
charge vanishes.
\end{abstract}

\noindent PACS number(s): 98.80.Cq, 98.80.Hw
\vfill
$^*$ This work was partially supported by the Swiss National Science
Foundation.
\end{titlepage}

\section{Introduction}
The developments in particle physics and cosmology
suggest that scalar fields may have played an important role in the
evolution of the early universe, for instance in primordial
phase transitions, and that they may constitute part of the dark
matter.
Moreover, scalar fields are an important ingredient in most of the
particle physics models
based on the unification of the fundamental forces, as for instance
in superstring
theories. Scalar particles are needed in cosmological
models based on inflation, whose relevance is supported by the
results of the COBE-DMR measurements that are consistent with
an Harrison-Zel'dovich spectrum
\cite{cob}.

These facts, in particular inflation,
motivated the study of the coupled Einstein-scalar field
equations to determine the time evolution and also
the gravitational equilibrium configurations of scalar fields,
which may form so-called boson stars \cite{kn:Jetzer:p1,kn:Liddle:p1}.

A detailed study of the solutions of the Einstein equations
for a homogeneous isotropic Friedmann-Lema\^{\i}tre universe with
a real scalar field has been carried out in particular
by Belinsky et al. \cite{bel1:p2,bel2:p2,bel3:p2}, Piran and Williams
\cite{kn:Piran:p2}.
The corresponding thorough investigation for a complex scalar field
has been performed in Ref.\cite{sci2:p2}

The knowledge of the behavior of the background solution in the early
universe is of importance for the analysis of the time evolution of
the perturbations \cite{kn:Deruelle:p2,kn:Mukhanov:p2}.
This is of relevance if scalar fields
make up part of the dark matter and form
compact objects, such as boson stars, or trigger the formation
of the observed large scale structures in the universe.

In this paper we study the time evolution of small perturbations in a gauge
invariant way for a complex
scalar field in the early zero curvature Friedmann-Lema\^{\i}tre
universe thus, generalizing the previous analysis
for a real scalar field \cite{mukp1p69}.
We consider a potential with a positive mass and a quartic self-interaction
term and 
closely follow the treatment developed by Mukhanov et al.
\cite{kn:Mukhanov:p2}. 
During the inflationary phase the behavior of
the perturbations is similar to the one of the real scalar field.
This is not surprising, since along the separatrices the phase of the
complex scalar field remains constant and thus inflation is
essentially driven by one component of the field \cite{sci2:p2}.
The short wavelength
perturbations get smeared out, whereas the long wavelength
perturbations increase.
On the other hand, the metric long wavelength perturbations are
entirely determined by 
the scale factor and hence survive the decay of the complex scalar
field background solution after inflation.

We extend our investigation of the
behavior of the perturbations also to the oscillatory phase, which
takes
place after inflation. Contrary to the real scalar field for which the
bosonic charge of the background solution vanishes, the charge
is different from zero in our case. 
If it is sufficiently large, we can
even find analytically the time evolution of the perturbations for a
massive complex scalar field. It turns out, that the long wavelength
perturbations decrease with time, whereas the short wavelength
perturbations, at best, oscillate around a constant value. Therefore, we do not
expect the formation of gravitational seeds around which structure can
form for the case of a massive complex scalar field with a
sufficiently
large bosonic charge. The other cases have to be treated
numerically.
In order to put constraints on the parameters of the scalar field
potential from the measured cosmic microwave background
(CMB) anisotropies \cite{cob}, we have to
analyze the quantized perturbations. 
Here, we have restricted the analysis only to the classical
perturbations.

Moreover, we also derive the Jeans wavenumber in the Newtonian regime
starting from the general relativistic equations, thus avoiding the
so-called ``Jeans swindle''. For the case where there is a
quartic self-interaction term in the potential the result we find
differs from the one given in Ref. \cite{khlop}.

The paper is organized as follows. In section \ref{section2:p3} we
present the basic equations, which we will use. In section
\ref{section3:p3} we give the derivation of the Jeans wavenumber
starting from the general relativistic equations. Section
\ref{section4:p3} is devoted to the analysis of the time evolution of
the
classical scalar mode fluctuations in a fully general relativistic
way, both in the inflationary and in the oscillatory phases. The
numerical results are presented in section \ref{section5:p3} and in
section \ref{section6:p3} we present briefly our conclusions. 

\section{Basic equations}\label{section2:p3}
We will consider the linear scalar mode perturbations since 
they are the only 
ones which contribute to the energy density fluctuations
\cite{kod}. 
As usual, we expand the scalar perturbations 
in terms of a complete set of harmonic functions 
$Y_k$, which are the eigenfunctions with eigenvalue $-k^2$ of the
Laplace-Beltrami operator 
$\mathbf{\Delta}$ defined on constant time slices $\Sigma_{\eta}$.   
In the following, we will omit for simplicity to write the 
subscript $k$ from $Y$.

The zero curvature
Fried\-mann\--Le\-ma\^{\i}\-tre metric---including the first order
perturbations---is then given by \cite{kod}
\begin{equation}
	ds^2=g_{\mu\nu}\theta^{\mu}\theta^{\nu}=-\theta^0\theta^0+
	\delta_{ij}\theta^i\theta^j, \label{metr}
\end{equation}
where 
\begin{gather*}
	\theta^0=\alpha d\eta=a\left(1+AY\right)d\eta \, ,\\ 
\begin{split}
	\theta^{i}&=\beta^{i}\,d\eta+a\left[\left(1+H_LY\right)\,dx^{i}
	+H_TY^{i}_{j}\,dx^{j}\right],\\
	&=-aBY^{i}\,d\eta+a\left[\left(1+H_LY\right)\,dx^{i}
	+H_TY^{i}_{j}\,dx^{j}\right],
\end{split}
\end{gather*}
and
\begin{displaymath}
	Y_i=\mbox{}-k^{-1}Y_{|i}\, , 
	\qquad Y_{ij}=k^{-2}Y_{|ij}+\frac{1}{3}Y\delta_{ij}\, .
\end{displaymath}
$\alpha$ is the lapse function and $\beta^{i}$ are the components of
the shift vector.
$A$, $B$, $H_L$ and $H_T$ are, respectively, the amplitude of the
perturbation of
the lapse function, the shift vector, the unit spatial volume and the
anisotropic distortion of $\Sigma_\eta$. All these functions, as well
as the scale factor $a$, depend only on the conformal time $\eta$.
The subscript
$\mbox{}_{|i}$ denotes the covariant derivative on $(\Sigma_\eta,\, 
\mathbf{\gamma})$, where $a^2\mathbf{\gamma}$ is defined as the
unperturbed induced 
metric on $\Sigma_\eta$. It is easy to verify that $Y_{ij}$ is traceless.

We take for the matter action
 \begin{equation}
  	S_m=\int \zeta \left[\frac{1}{2}g^{\mu\nu}\left(e_{\mu}\Phi 
  	e_{\nu}\Phi^{\ast}+e_{\nu}\Phi e_{\mu}\Phi^{\ast}\right)+
  	V\left(\Phi,\Phi^{\ast}\right)\right],
 	\label{smat}
 	 \end{equation}
where $\zeta$ is the volume form and $e_{\mu}$ is the dual basis of
$\theta^{\mu}$.
We expand the complex scalar field $\Phi$ to first order, where
$\phi(\eta)$ is the zero-order background solution and $\delta \phi
(\eta)Y$ 
is the first order term. The potential $V$ is such that $S_m$ is invariant
under a global $U(1)$ symmetry. Thus, V has to fulfill the following
conditions
\begin{eqnarray}
       \phi\frac{\partial V}{\partial \phi}-\phi^{\ast}
	\frac{\partial V}{\partial \phi^{\ast}} & = & 0 \, , 
	\label{v1}  \\
	\phi\frac{\partial V}{\partial \phi}+
	\phi^2\frac{\partial^2 V}{\partial \phi^2}-\phi\phi^{\ast}
	\frac{\partial^2 V}{\partial \phi \partial \phi^{\ast}} & = &
	0\, ,
	\label{v2}  \\
	\phi^{\ast}\frac{\partial V}{\partial \phi^{\ast}}+\phi^{\ast 2}
	\frac{\partial^2 V}{\partial \phi^{\ast 2}}-\phi\phi^{\ast}
	\frac{\partial^2 V}{\partial \phi \partial \phi^{\ast}} &= &
	0\, .
 	\label{v3}
\end{eqnarray}
The $U(1)$-global symmetry of $S_m$ implies a conserved current 
\begin{equation}
 	J=J^{(0)}+\delta J\, ,
 		\label{j}
\end{equation}
where
\begin{eqnarray}
	J^{(0)} & = & -\frac{i}{a}\left(\dot{\phi}^{\ast}\phi-
        \dot{\phi}\phi^{\ast}\right)e_0\, ,
	\label{j0}  \\
	\delta J & = & \left\{-\frac{i}{a}
	\left[\dot{\phi}^{\ast}\delta\phi-
	\dot{\phi}\delta\phi^{\ast}+\phi\dot{\delta\phi}^{\ast}-
	\phi^{\ast}\dot{\delta\phi}\right]-J^{(0)}A\right\}Ye_0
\nonumber \\ 
        &  & \mbox{}-\frac{ik}{a}\left[\phi\delta\phi^{\ast}-\phi^{\ast}
        \delta\phi\right]Y^i e_i\, .
	\label{dj}
\end{eqnarray}
Dot means the derivative with respect to $\eta$.
By varying $S_m$ with respect to the metric, we obtain the
energy-momentum  
tensor up to the first order terms
\begin{equation}
	T=\varepsilon e_0 \otimes e_0 + e_0 \otimes \mathbf{S}+
	\mathbf{S}\otimes e_0+\mathbf{T}
	\equiv T^{(0)}+\delta T,
	\label{tei}
\end{equation}
where	
\begin{eqnarray}
	\varepsilon &\equiv& \varepsilon^{(0)}+\delta \varepsilon
\nonumber 
	\\
	& = &\frac{1}{a^2}\biggl[\dot{\phi}^{\ast}\dot{\phi}+
	a^2 V \nonumber \\
	& &\mbox {}+\dot{\phi}^{\ast}\dot{\delta\phi}+
	\dot{\phi}\dot{\delta
\phi}^{\ast}-2AY\dot{\phi}^{\ast}\dot{\phi}+ 
	a^2\frac{\partial V}{\partial \phi}\delta \phi+
	a^2\frac{\partial V}{\partial \phi^{\ast}}
	\delta \phi^{\ast}\biggr] , 
	\label{t00} 
\end{eqnarray}
\begin{eqnarray}
        \mathbf{S} &\equiv& \mathbf{S}^{(0)}+
	\delta \mathbf{S}=\frac{k}{a^2}\left(\dot{\phi}\delta
\phi^{\ast}+ 
	\dot{\phi}^{\ast}\delta \phi \right) Y^i e_i\, ,
	\label{t0i} \\
	\mathbf{T}&\equiv& \mathbf{T}^{(0)}+\delta \mathbf{T} \nonumber
	\\
	& = & \frac{1}{a^2} \Biggl[ \dot{\phi} \dot{\phi}^{\ast}
	-a^2V+\biggl( \dot{\phi} \dot{\delta \phi}^{\ast} 
	+\dot{\phi}^{\ast}\dot{\delta\phi}-2A\dot{\phi}\dot{\phi}^{\ast}
	\nonumber\\
	& &\mbox{}
	-a^2\frac{\partial V}{\partial \phi}\delta \phi
	-a^2\frac{\partial V}{\partial \phi^{\ast}}\delta 
	\phi^{\ast}\biggr)Y\Biggr]\delta^{ij} e_i \otimes e_j\, .
	\label{tij} 
\end{eqnarray} 
The background quantities and the first order perturbation 
components can be easily identified.
To derive the Einstein equation using the $3+1$ formalism \cite{mis}, we 
first have to compute the second fundamental form $\mathbf{K}$ for the
slices  
$\Sigma_\eta$, for which we get
\begin{equation}
	\mathbf{K}=-\frac{1}{a}\left[h\delta^{ij}+k\sigma_g Y^{ij}+
	h\kappa_g Y\delta^{ij}\right]e_i \otimes e_j\, ,
	\label{2ff}
\end{equation}
where $h=\frac{\dot{a}}{a}$ is the Hubble expansion rate,
\begin{eqnarray*}
	\sigma_g & = & \frac{\dot{H}_T}{k}-B  
\end{eqnarray*}
and
\begin{eqnarray*}
	\kappa_g & = & -A+\frac{kB}{3h}+\frac{\dot{H}_L}{h}\, .
\end{eqnarray*}
The Einstein equations are then given by \cite{sta}
\begin{gather}
 	\mathbf{R}+[\mbox{Tr}(\mathbf{K})]^2-\mbox{Tr}(\mathbf{K}^2) =
 	16\pi G\varepsilon\, ,
	\label{g00}  \\
	\mathbf{\nabla}\cdot\mathbf{K}-\mathbf{\nabla}\mbox{Tr}\mathbf{K} =      
	8\pi G \mathbf{S}\, , 
	\label{g0i}  \\
 \begin{split}
	\partial_{t} \mathbf{K} = &\mathbf{L_{\beta}K}-\mathbf{Hess}
	(\alpha)+\alpha \bigl[\mathbf{Ric}(\mathbf{g})+
	2\mathbf{K}^2\\
	&\mbox{}+\mathbf{K}\mbox{Tr}\mathbf{K}
	-8\pi G\mathbf{T}-4\pi G\mathbf{g}(\varepsilon-\mbox{Tr}
	\mathbf{T})\bigr]\, , 
 \end{split}
	\label{gij}
\end{gather}
where Tr denotes the trace, $\mathbf{Hess}$ the Hessian, $\mathbf{g}$
the  
induced metric on $\Sigma_\eta$,  
$\mathbf{R}$ and $\mathbf{Ric}(\mathbf{g})$ are,
respectively, the Ricci scalar and the Ricci 
tensor of the slices $\Sigma_\eta$. The background
solution trivially satisfies eq.(\ref{g0i}), whereas from
eq.(\ref{g00}) and eq.(\ref{gij}) we obtain
\begin{eqnarray}
	3h^2 & = & 8\pi G\left[\dot{\phi}\dot{\phi}^{\ast}+a^2V\right],
	\label{e00}  \\
	\dot{h}+2h^2 & = & 8\pi Ga^2V.
	\label{eij}
\end{eqnarray}
>From eq.(\ref{gij}) we get for the terms in first order
\begin{equation}
 \begin{split}
	k^2Y_{ij}(\Phi_g+\psi)=&-Y\delta_{ij}
	\Biggl[-\ddot{\psi}-5h\dot{\psi}
 	-\frac{4}{3}k^2\psi \\
	&\mbox{}+h\dot{\Phi}_g+\left(16\pi Ga^2V-\frac{k^2}{3}\right)
	\Phi_g\\
	&\mbox{}+8\pi Ga^2\left(\frac{\partial
V}{\partial\phi}\delta\varphi+ 
	\frac{\partial V}{\partial\phi^{\ast}}\delta\varphi^{\ast}\right)
	\Biggr],
	\label{giij}
 \end{split}
\end{equation}
where we have introduced the gauge invariant quantities 
\cite{bar,kod,kn:Mukhanov:p2}
\begin{eqnarray}
	\psi & = & H_L+\frac{1}{3}H_T-\frac{1}{k}h\sigma_g\, ,
	\label{psi}  \\
	\delta\varphi & = & \delta\phi -\dot{\phi}\frac{\sigma_g}{k}\,
	,
	\label{varphi}  \\
	\Phi_g & = & A-\frac{1}{ka}(a\sigma_g)\spdot \, .
	\label{Phi}
\end{eqnarray}
For $i\neq j$ we obtain
\begin{equation}
	\psi=-\Phi_g\, .
	\label{eg}
\end{equation}
The meaning of the gauge invariant quantities is obvious 
on small
scales (i.e. $k\gg 1$), in which case they reduce to
$H_L+\frac{1}{3}H_T$ (the  
intrinsic scalar curvature perturbation), 
$\delta\phi$ (the complex scalar field perturbation) 
and $A$ (the gravitational 
potential), respectively.
 
The first order terms of eqs.(\ref{g00})-(\ref{g0i}) combined with 
the trace of\\eq.(\ref{giij}) lead to the following Einstein equations
\begin{eqnarray}
	6h\dot{\psi}+\left(2k^2+6h^2\right)\psi &= & 8\pi G
\biggl(\dot{\phi}^{\ast}\dot{\delta\varphi}+ 
	\dot{\phi}\dot{\delta\varphi}^{\ast}
	+2\dot{\phi}\dot{\phi}^{\ast}\psi \nonumber\\
	&  & \mbox{}+a^2\frac{\partial V}{\partial\phi}\delta\varphi+
	a^2\frac{\partial
V}{\partial\phi^{\ast}}\delta\varphi^{\ast}\biggr), 
	\label{de00}   \\
	\dot{\psi}+h\psi & = & -4\pi
G\left(\dot{\phi}^{\ast}\delta\varphi+ 
	\dot{\phi}\delta\varphi^{\ast}\right),
	\label{de0i} \\
	\ddot{\psi}+6h\dot{\psi}+\left(k^2+16\pi Ga^2 V\right)\psi & = &
        8\pi Ga^2\left(\frac{\partial V}{\partial\phi}\delta\varphi+
        \frac{\partial
V}{\partial\phi^{\ast}}\delta\varphi^{\ast}\right), 
 	\label{deij}
\end{eqnarray}
where we used eq.(\ref{eg}).

By varying $S_m$ with respect to $\Phi^{\ast}$
we obtain the Klein-Gordon equation, which to the zero order is given
by
\begin{equation}
	\ddot{\phi}+2h\dot{\phi}+a^2\frac{\partial V}{\partial \phi^{\ast}}=0.
	\label{kg}
\end{equation}
By directly expressing the first order perturbed Klein-Gordon equation in 
terms of the above defined gauge invariant quantities, we get
\begin{equation}
	\ddot{\delta\varphi}+2h\dot{\delta\varphi}+k^2\delta\varphi+
	4\dot{\phi}\dot{\psi}-
	2a^2\frac{\partial V}{\partial\phi^{\ast}}\psi+
	a^2\frac{\partial^2 V}{\partial\phi^{\ast 2}}\delta\varphi^{\ast}+
	a^2\frac{\partial^2 V}{\partial\phi^{\ast}\partial\phi}\delta\varphi=0.
	\label{dkg}
\end{equation}
Similarly by varying $S_m$ with respect to $\Phi$ we obtain the complex
conjugate of eqs.(\ref{kg})-(\ref{dkg}).
It should be noticed that the first order terms are entirely determined by
eqs.(\ref{de00})-(\ref{de0i}), eq.(\ref{dkg}) and its complex conjugate.
These equations hold for all values of $k$ 
and, therefore, the range of validity is
not limited by the horizon scale.
The solutions of these equations will be valid as long as 
$\left|\delta\varphi/\phi\right|$ and $\psi$ are sufficiently small,
irrespective of $k$.
 
The background solution is also 
established by eq.(\ref{e00}), eq.(\ref{kg}) and its complex conjugate.
The fact that eq.(\ref{eij}) and eq.(\ref{deij}) are automatically
satisfied is due to the Bianchi identities.

The conservation of the current $J$ leads to a set of two additional
equations. For the 
background part of the solution, we get the conservation with respect to the
conformal time of the bosonic charge
\begin{equation}
\Xi=\frac{i\, a^2}{2}\left(\dot{\phi}^{\ast}\phi-
\dot{\phi}\phi^{\ast}\right)\, ,
\label{Xi}
\end{equation}
and for the first order term the expression
\begin{equation}
	\partial_\eta \xi=a^2k^2\left(\phi^{\ast}\delta\varphi-
	\phi\delta\varphi^{\ast}\right),
	\label{ddj}
\end{equation}
where
\begin{equation}
	\xi=4a^2\psi\left(\dot{\phi}^{\ast}\phi-\dot{\phi}\phi^{\ast}\right)+a^2
	\left(\dot{\phi}^{\ast}\delta\varphi
	-\dot{\phi}\delta\varphi^{\ast}+\phi\dot{\delta\varphi}^{\ast}-
	\phi^{\ast} \dot{\delta\varphi}\right).
	\label{xi}
\end{equation}
For very large wavelengths $\xi$ is also a conserved quantity, since 
then the right hand side of eq.(\ref{ddj}) vanishes up to first
order.

\section{The Newtonian regime}\label{section3:p3}
In this section, we derive the Jeans wavenumber starting from the 
general relativistic equations. This way, we avoid the so-called ``Jeans 
swindle'' \cite{wei}. The gauge invariant formalism is very well adapted 
to this problem, since in the longitudinal mode ($B=E=0$) $\psi$ is just 
the Newtonian potential. 

To obtain the Jeans wavenumber we have to derive the 
dispersion relation. We consider a potential with a mass and a 
quartic self-interaction term. The Newtonian regime is 
realized when the expansion of the universe can be neglected and
the horizon size is much larger than the wavelength of the perturbation
(i.e. $k\gg h$ and $h\sim \epsilon$, where $\epsilon\ll 1$).
With these conditions it follows from eq.(\ref{e00}) that 
\begin{align}
k^2\gg &8\pi G \dot{\phi}\dot{\phi}^{\ast}, 
\label{approx1} \\
\intertext{and}
k^2\gg &8\pi G a^2V\equiv 8\pi G
a^2\left[m^2\phi\phi^{\ast}+\lambda(\phi\phi^{\ast})^2\right]. 
\label{approx2}
\end{align}
We obtain two conditions from eq.(\ref{e00}), because the right hand side of
this equation is a sum of two positive terms.
>From eq.(\ref{eij}) we have $\dot{h}\ll k^2$ and thus
$\left|\ddot{a}/a\right|\ll k^2$.
We have now to find the solution for the background in the Newtonian
regime. Inserting
\begin{equation}
\phi=f/a \label{defp}
\end{equation}
into eq.(\ref{kg}) we get
\begin{equation}
\ddot{f}+f\left[a^2m^2+2\lambda ff^{\ast}-\dot{h}-h^2\right]=0.
\label{nkg}
\end{equation}
Imposing 
\begin{equation}
a^2m^2+2\lambda ff^{\ast}\gg\left|\dot{h}+h^2\right|=\left|\ddot{a}/a\right|
\label{approx3}
\end{equation}
---which means, that for an oscillation period of $\phi$,
$a$ can be considered as a constant---
eq.(\ref{nkg}) can be solved and we get
\begin{gather}
f=f_0\exp (i\omega \eta)\, ,\\
\intertext{where}
\omega=\sqrt{a^2m^2+2\lambda f_0^2}\, ,
\label{defw}
\end{gather}
and $f_0$ is an integration constant.
Using eq.(\ref{defp}), we have
\begin{equation}
\phi=b_0\exp (i\omega \eta)
\label{nsp}
\end{equation}
with $b_0=f_0 /a$, which within our approximations is constant. 
Hence,\\eqs.(\ref{approx1})-(\ref{approx2}) and eq.(\ref{approx3})
can be rewritten as
\begin{align}
k^2 &\gg 8\pi Gb_0^2a^2\left(m^2+2\lambda b_0^2\right), \label{approxa} \\
k^2 &\gg 8\pi Gb_0^2a^2\left(m^2+\lambda b_0^2\right), \label{approxb}  \\
\left|\frac{\ddot{a}}{a}\right| &\ll a^2m^2+2\lambda a^2b_0^2.
\label{approxc}
\end{align}
To obtain the dispersion relation we perform the following expansion
\begin{gather}
\Phi\equiv\left(b_0+\frac{b_1Y}{a}\right)\exp \left(i\omega
\eta+i\frac{\varphi Y}{a}\right)=\phi+\frac{Y}{a}\left(b_1+ib_0\varphi\right)
e^{i\omega \eta}+\mathcal{O}(2)\, ,
\label{defPhi} \\
\psi = \frac{\Psi}{a}\, . \label{defpsi}
\end{gather}
As next, we rewrite eq.(\ref{de00}) and eq.(\ref{dkg}) in terms of
$\Psi$, $b_1$ and $\varphi$. Eq.(\ref{dkg}) will then split into a real
and an imaginary part. With the following expansion
\begin{equation}
b_1=b_{10}\exp \left(\Omega \eta\right)\, , \qquad
\varphi=\varphi_{10}\exp \left(\Omega \eta\right)\, , \qquad
\Psi=\Psi_{10}\exp \left(\Omega \eta\right) \label{psi10}
\end{equation}
and using the above mentioned Newtonian approximations, we get a set of 3
equations
\begin{gather}
16\pi Ga^2b_0b_{10}\left(m^2+2\lambda b_0^2\right)+8\pi
Gb_0^2\Omega\omega\varphi_{10}-k^2\Psi_{10}=0\, , \label{nde00} \\
b_{10}\left(\Omega^2+k^2+4\lambda a^2b_0^2\right)-2\Omega\omega
b_0 \varphi_{10}-2\Psi_{10}a^2b_0\left(m^2+2\lambda b_0^2\right)=0\, , 
\label{rndkg}
\\
2\omega\Omega b_{10}+\varphi_{10}b_0\left(\Omega^2+k^2
\right)+4\Psi_{10}b_0\omega\Omega=0\, . \label{indkg} 
\end{gather}
After some algebraic manipulations, we find
\begin{gather}
\begin{bmatrix}
16\pi Gb_0\omega^2 & 8\pi Gb_0^2\omega\Omega & -k^2 \\
0 & \mbox{}-\frac{\Omega b_0}{2\omega}\left(4\omega^2+\Omega^2+k^2+4\lambda
a^2b_0^2\right) & a_{22} \\
0 & 0 & a_{33}
\end{bmatrix}
\begin{bmatrix}
b_{10} \\
\varphi_{10} \\
\Psi_{10}
\end{bmatrix}
=0\, , \label{matrix} \\
\intertext{with}
a_{22}=\frac{1}{16\pi Gb_0\omega^2}\left(-32\pi Gb_0\omega^4 
+k^2\Omega^2+k^4+4\lambda a^2b_0^2k^2\right) \label{defa22} \\
\intertext{and}
\begin{split}
a_{33}=& \frac{1}{32\pi Gb_0a^2\omega^2k^2} \Bigl[ -64\pi
Gb_0^2\omega^4k^2+2k^6+8\lambda a^2b_0^2k^4 \\
&\mbox{}+\Omega^2 \left(2k^4+64\pi Gb_0^2\omega^2+2k^2\right)
\left(4\omega^2+\Omega^2
+k^2+4\lambda a^2b_0^2\right)\Bigr]\, . 
\end{split} \label{disp}
\end{gather}
$a_{33}=0$ is the dispersion relation, which
is the required condition for having a non-trivial
solution of eq.(\ref{matrix}). The Jeans wavenumber, corresponding to
the solution of the dispersion relation with $\Omega=0$, is given by
\begin{equation}
k^2_j= 2a^2b_0\left[\sqrt{\lambda^2b_0^2+8\pi G(m^2+2\lambda b_0^2)^2}
-\lambda b_0\right]. \label{jeans}
\end{equation}
For $\lambda=0$ we obtain the same result as in Ref.\cite{khlop}.
Notice that, since we have taken the conformal time, we have an
over-all $a^2$ factor.
For $\lambda\neq 0$, we disagree with the expression for the Jeans wavenumber
given\footnote{\hskip 0.2cm $k^2_j\simeq \frac{8\pi
Gm^2(m^2+\left|\lambda\right|^2a^2)}{\left|\lambda\right|^2}$} by eq.(31)
in Ref.\cite{khlop}.
This is due to the fact that eq.(1)
in Ref.\cite{khlop} is not compatible with a
quartic self-interaction
term. Indeed, for the mass-less complex scalar field with $\lambda\neq 0$, we
expect to have a non-vanishing Jeans wavenumber as
it is the case for our result, given by eq.(\ref{jeans}), but not for
the corresponding eq.(31) in Ref.\cite{khlop}.
Only disturbances whose wavenumber is smaller than 
the Jeans wavenumber 
$k_j$ can grow. Hence, the knowledge of the Jeans wavelength 
$\lambda_j=2\pi/k_j$ gives a rough idea of the size of the objects 
which can be formed by gravitational collapse. What we do not know, however, 
is the rate at which the perturbations will grow. 

\section{Classical perturbations of a complex scalar field}\label{section4:p3}

In this section we analyze the time evolution of the scalar mode
fluctuations in a fully general relativistic way. We will consider the
long wavelength and the short wavelength limits separately. To solve the
Einstein eqs.(\ref{de00})-(\ref{de0i}), we first define the complex
valued function 
$U(\eta)$ as the solution of the following differential
equation
\begin{equation}
\dot{U}+hU=-4\pi G\dot{\phi}^{\ast}\delta\varphi~.\label{a1}
\end{equation}
For \hbox{$\left|\delta\varphi\right|\lower 4pt \vbox{\vskip
3pt\hbox{$<$}\vskip -8.82pt \hbox{$\sim$}} \hskip 4pt
|\dot{\phi}^{\ast}|/ 4\pi G$} 
the complex scalar field perturbation is solution of the constraint 
eqs.(\ref{de00})-(\ref{de0i}) if and only if the Bardeen potential
$\psi$ and $U$ satisfy the following equations
\begin{gather}
\psi=U+U^{\ast}+\frac{k_{\psi}}{a}~,\label{apsi}\\ 
\ddot{U}+2\left(h-\frac{\ddot{\phi}^{\ast}}{\dot{\phi}^{\ast}}\right)\dot{U}
+\left(k^2+2\dot{h}-2h\frac{\ddot{\phi}^{\ast}}{\dot{\phi}^{\ast}}\right)U=
-\frac{k_{\psi}}{2a}\left(k^2+\dot{h}-h^2\right)~,
\label{a2}
\end{gather}
the r.h.s of the last equation represents a source term and $k_\psi$ is a
real integration constant. In fact the term $k_\psi /a$ in
eq.(\ref{apsi}) is a solution of the corresponding homogeneous
eq.(\ref{de0i}), namely $\dot{\psi}+h\psi=0$. In the real scalar
field case there is no source term \cite{kn:Mukhanov:p2}.
To simplify the notation
we omit to write the explicit $k$ dependence of $U$.
Setting $U=\frac{\dot{\phi}^{\ast}u}{a}$, eq.(\ref{a2}) can be rewritten as
\begin{equation}
\ddot{u}+k^2u+\left(\dot{h}-h^2+
\frac{\dddot{\phi^{\ast}}}{\dot{\phi}^{\ast}}
-2\frac{\ddot{\phi}^{\ast 2}}{\dot{\phi}^{\ast 2}}\right)u=
-\frac{k_{\psi}}{2\dot{\phi}^{\ast}}\left(k^2+\dot{h}-h^2\right)~.
\label{u2}
\end{equation}
With $g=\frac{h}{a\dot{\phi}^{\ast}}$ and using the background
equation we find 
\begin{equation}
\ddot{u}+k^2u+\left(\frac{32\pi G i\Xi}{a^2}
-\frac{\ddot{g}}{g}\right)u=
-\frac{k_{\psi}}{2\dot{\phi}^{\ast}}\left(k^2+\dot{h}-h^2\right)~.
\label{u22}
\end{equation}
By solving this last equation
we get immediately $U$. Using eq.(\ref{apsi}) and eq.(\ref{a1}) 
we obtain $\psi$ and $\delta\varphi$, respectively.

In the analysis of eq.(\ref{u22}) we will restrict ourselves 
to long and short wavelength perturbations
in the inflationary and the oscillatory phase of the background
solution.

\subsection{Perturbations during inflation}\label{section4.1:p3}

We consider inflation generated by the
potential 
\begin{equation}
V=m^{2}\phi \phi^{\ast}+\lambda
(\phi \phi^{\ast})^{2}\, ,
\label{potential}
\end{equation}
for which during inflation
$\left|\phi\right|^2\dot{\vartheta}$ is asymptotically 
zero \cite{sci2:p2}, where we have set $\phi = |\phi| e^{i\vartheta}$.
>From eq.(\ref{Xi}) we get 
\begin{equation}
a^2{\left|\phi\right|}^2\dot{\vartheta}=\Xi, \label{Xit}
\end{equation}
where $\Xi$ is the constant bosonic charge. Hence, inflation must
start with a small value of $\Xi / a^2$.
On the separatrix, where inflation
occurs,  $\Xi / a^2$ decays exponentially, whatever the values of $m$
and $\lambda$ are. As a consequence eq.(\ref{u22}) 
reduces to
\begin{equation}
\ddot{u}+k^2u-\frac{\ddot{g}}{g}u=
-\frac{k_{\psi}}{2\dot{\phi}^{\ast}}\left(k^2+\dot{h}-h^2\right)~.
\label{u22i}
\end{equation}
Moreover, we see from eq.(\ref{Xit}) that for small values of
$\Xi$ the inflationary phase lasts longer.
Eq.(\ref{u22i}) is actually valid not only for the inflationary stage
but also when $\Xi /a^2$ is negligible and \hbox{$|\dot{\phi}^{\ast}|
\lower 4pt \vbox{\vskip 3pt\hbox{$>$}\vskip -8.82pt \hbox{$\sim$}}
4\pi G\left|\delta\varphi\right|$}.

During inflation we have for $\lambda\neq 0$ and
$\lambda =0$, respectively 
\begin{align}
\ddot{g}/g &\simeq \frac{a^2V_{\phi\phi^{\ast}}}{2}
\simeq -\frac{3}{2}a^2H_t = -\frac{3}{2}\left(\dot{h}-h^2\right)~,\\
\intertext{and} 
\ddot{g}/g &\simeq \frac{2}{3}a^2V_{\phi\phi^{\ast}}\simeq
-2a^2H_t = -2 \left(\dot{h}-h^2\right)~,
\end{align}
where we have denoted
$\frac{\partial V}{\partial\phi\partial\phi^{\ast}}$ by
$V_{\phi\phi^{\ast}}$ and introduced
$H=\frac{1}{a}\frac{da}{dt}\equiv \frac{a_t}{a}$ with $t$ being the
real time defined as $dt=a\,d\eta$ ($H_t$ means derivative of $H$ with
respect to real time).
This result can
be easily obtained using the inflationary background solutions (see
\cite{sci2:p2}). 
Since inflation occurs only when $V \gg \phi_t\phi_t^\ast$,  
we have $\left|H_t\right|\ll H^2$.
This means that $\ddot{g}/g \ll a^2H^2$ and $\ddot{g}/g$ increases
exponentially. Hence the time interval, where
$k\simeq \ddot{g}/g$, is very short.

For  $k\ll \ddot{g}/g$ --- the long wavelength perturbations ---  
the integration of eq.(\ref{u22i}) gives
\begin{equation}
u=k_1\;g+k_2\;g\int \frac{d\eta}{g^2}-g\int
\frac{1}{g^2}\left[\int^{\eta}
\frac{k_{\psi}\left(\dot{h}-h^2\right)}{2\dot{\phi}^{\ast}}
d\tilde{\eta}\right]\;d\eta \label{ukl1} 
\end{equation}
with $k_1$ and $k_2$ being complex integration constants. Using the
background equations and the fact that the phase of the complex scalar
field is constant, we can rewrite eq.(\ref{ukl1}) as
\begin{equation}
u=\frac{k_2e^{-2i\vartheta}}{8\pi
G\dot{\phi}^\ast}\frac{d}{d\eta}\left[\frac{1}{a}\int^\eta 
a^2d\tilde{\eta}\right]-\frac{k_{\psi}}{2\dot{\phi}^{\ast}}~. \label{ukl2}
\end{equation}
In this last expression the second integration constant has been absorbed in
the integral. It follows that for this case the second term in
eq.(\ref{ukl2}), which is due to the source term (see eq.(\ref{a2})),
gives no contribution as we will see below.
With eq.(\ref{ukl2}) we obtain for the gauge invariant metric
perturbation
\begin{align}
\begin{split}
\psi &=\frac{2}{a} Re\left(\tilde{k_2}\right)\left[\frac{1}{a}\int^\eta
a^2d\tilde{\eta}\right]\spdot 
= 2 Re\left(\tilde{k_2}\right)\frac{d}{d\,t}\left[\frac{1}{a}\int^t
a\,d\tilde{t}\right]\\
&= 2 Re\left(\tilde{k_2}\right)\left[1-\frac{H}{a}\int^ta\,d\tilde{t}\right],\\
&= 2 Re\left(\tilde{k_2}\right)\left(\left[H^{-1}\right]\spdot
   -\left[H^{-1}\left[H^{-1}\right]\spdot\right]\spdot
   +\left[H^{-1}\left[H^{-1}\left[H^{-1}\right]\spdot\right]
   \spdot\right]\spdot-\dotsb\right)\label{plw}
\end{split}
\end{align}
and during the inflationary stage
\begin{equation}
\psi \simeq -2 Re\left(\tilde{k}\right)\frac{H_t}{H^2},
\end{equation}
where $\tilde{k_2}=k_2e^{-2i\vartheta}/8\pi G$ and $Re$ means the real part.
For the gauge invariant field perturbation we get 
\begin{align}
\delta\varphi &=-\frac{k_2\dot{\phi}^\ast}{4\pi G\,a^2}\int^\eta
a^2\,d\tilde{\eta} = -\frac{k_2\phi_t^\ast}{4\pi G\,a}\int^ta\,d\tilde{t}\\
\intertext{and during inflation}
\delta\varphi &\simeq -\frac{k_2\phi_t^\ast}{4\pi G\,H}.
\end{align}

For $k\gg \ddot{g}/g$ --- the short wavelength perturbations --- 
eq.(\ref{u22i}) can be integrated and we obtain
\begin{equation}
\begin{split}
u=\gamma_1 & \cos k\eta + \gamma_2\sin k\eta + \frac{\cos
k\eta}{k}\int^\eta \frac{k_\psi 
k^2}{2\dot{\phi}^\ast}\sin k\tilde{\eta}\; d\tilde{\eta} \\
&\hbox{\hskip -3cm}- \frac{\sin k\eta}{k}\int^\eta \frac{k_\psi
k^2}{2\dot{\phi}^\ast}\cos k\tilde{\eta}\; d\tilde{\eta}~.
\end{split}\label{uswli}
\end{equation}
As a consequence,
\begin{align}
\psi &=2 Re\left[\frac{\dot{\phi}}{a}\left(\gamma_1\cos
(k\eta)+\gamma_2\sin (k\eta)\right)\right]+\frac{k_{\psi}}{a}\left(1-
\frac{3k^2}{4M_p^2\lambda+3k^2}\right)~,\nonumber \\
&\simeq 2
Re\left[\phi_t\left(\gamma_1\cos\left(k\int^t\frac{d\tilde{t}}{a}\right)+ 
\gamma_2\sin \left(k\int^t\frac{d\tilde{t}}{a}\right)\right)\right],\label{psisw}
\end{align}
\begin{align}
\begin{split}
\delta\varphi &=-\frac{1}{4\pi G}\Bigg\{\frac{k}{a}
\left[\gamma_2^\ast\cos (k\eta)-\gamma_1^\ast\sin
(k\eta)\right]\\
&\quad+\frac{\ddot{\phi}^\ast}{a\dot{\phi}^\ast}\left[\gamma_1^\ast\cos
(k\eta)+\gamma_2^\ast\sin(k\eta)\right]\Bigg\},
\end{split}\label{varphisw}
\end{align}
where $\gamma_1, \gamma_2$ are
complex integration constants and $M_p^2=1/8\pi G$. Notice, that 
also for this case the term proportional to $k_\psi$ does practically
not affect $\psi$. During
inflation for a perturbation 
wavelength smaller than the Hubble ra\-dius --- $k^2\gg a^2H^2\simeq
\ddot{\phi}^\ast /2\dot{\phi}^\ast$ --- we find 
\begin{equation}
\delta\varphi \simeq -\frac{k}{4\pi G\,a}
\left[\gamma_2^\ast\cos (k\eta)-\gamma_1^\ast\sin
(k\eta)\right].\label{phisw}
\end{equation}
The slow-rolling approximation,
required for having a sufficiently long
inflationary stage, leads to a slowly variation of the mean value of
$\psi$.
The behavior of $\delta\varphi$ is governed by the $1/a$
factor, which decreases rapidly. It follows, as expected, that the
short wavelength fluctuations of the scalar field are smeared out.


We consider an initial perturbation with wavelength smaller than the
Hubble radius 
($k^2\gg a^2H^2\gg \ddot{g}/g$), which will be outside it at the end of
inflation ($k^2\ll a^2H^2$). Indeed there are wavelengths that fulfill
these conditions, since $a^2H^2$ increases exponentially
(see Fig. 1).
At the end of inflation, the evolution of the gauge invariant metric
potential is given by eq.(\ref{plw}). Later on, when the universe is
dominated by relativistic particles, the scale factor is a power-law
function
$a\propto t^\mu$. Hence, the Hubble radius increases more rapidly than
the fixed comoving wavelength. The metric perturbation can thus re-enter
inside the Hubble radius and induce fluctuations on the ordinary
matter. These fluctuations would be the source for the
anisotropies in the CMB, which have been
measured by COBE \cite{cob}.

At the Hubble radius crossing, with eq.(\ref{plw}), the metric
perturbation is given by
\begin{equation}
\psi=2 Re\left(\tilde{k_2}\right)\frac{1}{\mu+1}.
\end{equation}
We can now use the relation between $\tilde{k}$ and $\delta\varphi$ at
the time $t_c$, defined such that
$k=\ddot{g}/g$, and substitute it into the last equation. We get
\begin{equation}
\psi=- Re\left(\frac{\delta\varphi}{\phi_t^\ast}H e^{-2i\vartheta}
\right)_{t=t_c}\frac{1}{\mu+1}.\label{psiout}
\end{equation}
There is an amplification of the metric perturbation between the time
it leaves the Hubble radius and the time it re-enters. Using the
asymptotic solutions for the inflationary stage this amplification
can be derived analytically. As an example for $\lambda = 0$ we
obtain
\begin{equation}
\frac{\psi_{in}}{\psi_{out}}=\frac{m^2t_c^2}{3\left(\mu+1\right)} 
~.
\end{equation}
This generalizes the result of previous investigations 
for the real scalar field case \cite{mukp1p69}. In
eq.(\ref{psiout}), the gravitational potential still depends on
the value of the complex scalar field perturbation at the time
$t=t_c$. In order to be able to put constraints 
on the parameters of the model, by comparing with the observed CMB
anisotropies,  
one should be able to express
$\psi$ as a function of the background quantities only. 
To achieve this one has, as for the real scalar case
\cite{Mk}, to quantize
the complex scalar field. However, this task is
not as simple as for the real scalar field, since 
in the lagrangian there is 
an additional interaction term between the real and the complex
part of the scalar field. 
We plan to come back on this issue in a future publication.

\subsection{Perturbations during the oscillatory
phase}\label{section4.2:p3} 

After the inflationary stage the background solution goes through an
oscillatory phase. This corresponds in the phase portrait to
winding around a focus point $A$ for which $\phi = \dot{\phi} =0$.
For a massive complex scalar field the asymptotic behavior of the
background solutions around the point $A$ is discussed in Ref.\cite{sci2:p2}.
Expressing the asymptotic behavior as function of the conformal time, 
we have
\begin{equation}
a=3a_0m^2 \eta^2\, , \label{aeta}
\end{equation}
\begin{equation}
\begin{split}
\phi=\frac{2M_p}{\sqrt{3}\, a_0m^3\eta^3}&\left[\cos\vartheta_{30}\cos
(a_0m^3\eta^3-\eta_1)\right.\\
&\left.\quad +i\sin\vartheta_{30}\cos (a_0m^3\eta^3-\eta_2)\right]\, ,
\end{split} \label{phiosc}
\end{equation}
\begin{equation}
\begin{split}
\dot{\phi}=-\frac{2\sqrt{3}\, M_p}{\eta}&\left[\cos\vartheta_{30}
\sin (a_0m^3\eta^3-\eta_1)\right.\\
&\left.\quad +i\sin\vartheta_{30}\sin (a_0m^3\eta^3-\eta_2)\right]\, ,
\end{split} \label{dotphiosc}
\end{equation}
for $\eta\rightarrow\infty$.
$a_0$, $\eta_1$, $\eta_2$, $\vartheta_{30}$ are dimensionless integration
constants. Due to
the following scaling behavior
\begin{align*}
a(a_0^{-1/3}\eta)&=a_0^{1/3}a(\eta)\, ,\\
\phi(a_0^{-1/3}\eta)&=\phi(\eta)\\
\intertext{and}
\dot{\phi}(a_0^{-1/3}\eta)&=a_0^{1/3}\dot{\phi}(\eta)
\end{align*}
of the background equations, we can restrict the analysis to $a_0=1$.

For the study of the first order fluctuations eq.(\ref{u2}) can
apriori only
be used as long as \hbox{$|\dot{\phi}^{\ast}|
\lower 4pt \vbox{\vskip 3pt\hbox{$>$}\vskip -8.82pt \hbox{$\sim$}}
4\pi G\left|\delta\varphi\right|$}.
$|\dot{\phi}^{\ast}|$ is a damped oscillating function which has its
minimum on an oscillation period when
\begin{equation}
b\equiv\cos^2\vartheta_{30}\sin^2(m^3\eta^3-\eta_1)+
\sin^2\vartheta_{30}\sin^2(m^3\eta^3-\eta_2)
\end{equation}
is minimum. If $b$ is large enough (such as $b/\eta>|\delta\varphi|$)
then eq.(\ref{u2}) will be valid for a large number of oscillations.
On the contrary, if the minimum of $b$ is very small,
then there will be for each oscillation period two time intervals
where eq.(\ref{u2}) is no longer expected to be valid.
One can easily see that the minimum
of $b$ vanishes for
$\vartheta_{30}=n\pi/2$ or $(\eta_2-\eta_1)=n\pi$, where $n$ is an
integer. These conditions correspond to a vanishing 
bosonic charge $\Xi$. Indeed, inserting
eqs.(\ref{aeta})-(\ref{dotphiosc}) in eq.(\ref{Xi}), we find
\begin{equation}
\Xi=2\, \Xi_{max}\cos\vartheta_{30}\sin\vartheta_{30}\sin (\eta_2-\eta_1)\, ,
\label{Xiosc}
\end{equation}
where $\Xi_{max}=18M_p^2m$ is the maximum bosonic charge.
The maximum value of all possible minima of $b$ is reached for
$\vartheta_{30}=\pi/4+n\pi/2$ and $ \eta_2-\eta_1=\pi/2+l\pi$
($l$ being an integer).
For these and only these values of $\vartheta_{30}$ and
$(\eta_2-\eta_1)$, $b=1/2$ and 
$|\Xi|$ is maximum. One can show that by increasing the minimum of $b$
also the absolute value of the bosonic charge increases.
Following these lines, we see that the time interval,
where eq.(\ref{u2}) is valid, increases with $|\Xi|$. Hence, we will
restrict our analysis to $\Xi\sim \Xi_{max}$.

With eqs.(\ref{aeta})-(\ref{dotphiosc}) we can easily show that
\begin{equation}
\left|\frac{\dddot{\phi}^\ast}{\dot{\phi}^\ast}-
2\frac{\ddot{\phi}^{\ast 2}}{\dot{\phi}^{\ast 2}}\right|\simeq 
\frac{9m^6\eta^4}{b}\left(\left(b-2\right)^2-
2\frac{\Xi^2}{\Xi_{max}^2}\right)\, .
\end{equation}
The right hand side can never be zero for any
value of $b$. It follows that 
\begin{equation}
|\dot{h}-h^2|\ll\left|\frac{\dddot{\phi}^\ast}{\dot{\phi}^\ast}-
2\frac{\ddot{\phi}^{\ast 2}}{\dot{\phi}^{\ast 2}}\right|\, .
\end{equation}
Thus, eq.(\ref{u2}) reduces to
\begin{equation}
\ddot{u}+k^2u-\frac{\ddot{p}}{p}u=
-\frac{k_{\psi}}{2\dot{\phi}^{\ast}}\left(k^2+\dot{h}-h^2\right)~,
\end{equation}
where $p=1/\dot{\phi}^\ast$. There are three regimes to study. The first
is given by the condition $k^2\gg |\ddot{p}/p|\gg |\dot{h}-h^2 |$, the second
by $|\ddot{p}/p|\gg k^2\gg |\dot{h}-h^2 |$ and the
third by $|\ddot{p}/p|\gg |\dot{h}-h^2 |\gg k^2$. 

For the short wavelength perturbations --- $k^2\gg|\ddot{p}/p|$ --- 
eq.(\ref{uswli}) can again be used. Using that for $\Xi\sim\Xi_{max}$
\begin{gather}
\phi\simeq \mp i\,\frac{\sqrt{2}M_p}{\sqrt{3}m^3\eta^3}e^{\pm
i\left(m^3\eta^3-\eta_0\right)}\label{phios}\\ \intertext{and}
\dot{\phi}\simeq \frac{\sqrt{6}M_p}{\eta}e^{\pm
i\left(m^3\eta^3-\eta_0\right)}\label{dphios}~,
\end{gather}
we get that
\begin{align}
\psi &\simeq 2 Re\left[\frac{\dot{\phi}}{a}\left(\gamma_1\cos
(k\eta)+\gamma_2\sin
(k\eta)\right)\right]+\frac{k_{\psi}}{a}+\mathcal{O}(1/\eta^4)~,\nonumber
\\ 
&\simeq 2 Re\left[\phi_t\left(\gamma_1\cos\left(k\int^t\frac{d\tilde{t}}{a}\right)+
\gamma_2\sin
\left(k\int^t\frac{d\tilde{t}}{a}\right)\right)\right]+\frac{k_{\psi}}{a},\label{ospsik2}
\end{align}
and $\delta\varphi$ is given by eq.(\ref{varphisw}), since it can
be shown that the inhomogeneous solution is of order
$1/\eta^6$.
The different signs in eqs.(\ref{phios})-(\ref{dphios}) depend on the
integers $n$ and $l$.
We see that the dominant term of $\psi$ decreases like $1/a$.
With eq.(\ref{varphisw}) we have
\begin{equation}
\delta\varphi \simeq-\frac{1}{4\pi G}
\frac{\ddot{\phi}^\ast}{a\dot{\phi}^\ast}\left[\gamma_1^\ast\cos
(k\eta)+\gamma_2^\ast\sin(k\eta)\right],
\end{equation}
since in general
\begin{equation}
\begin{split}
\frac{\ddot{\phi}^\ast}{a\dot{\phi}^\ast}=\frac{m}{2b}
&\left[\cos^2\vartheta_{30}
\sin (2m^3\eta^3-2\eta_1)\right.\\
&\quad \left.+\sin^2\vartheta_{30} \sin (2m^3\eta^3-2\eta_2)
-i\frac{\Xi}{\Xi_{max}}\right]
\end{split}
\end{equation}
and $k/a$ is a decreasing function. It follows that 
the real and the imaginary part of $\delta\varphi$
oscillate around a constant value. 

For the long wavelength case --- $k^2\ll|\dot{h}-h^2|\ll|\ddot{p}/p|$
--- $u$ is
now given by eq.(\ref{ukl1}), where we have to substitute $g$ by
$p$. It is easy to see that for large~$t$
\begin{equation}
\psi=\frac{Re\left(k_1\right)}{a}+\frac{1}{t}\left(k_3+k_3^{\ast}\right)+
\mathcal{O}\left(\frac{1}{t^2}\right), \label{psilwos}
\end{equation}
where 
\begin{equation}
k_3=2k_2M_p^2\left(\sin^2\vartheta_{30}-\cos^2\vartheta_{30}
    +2i\sin\vartheta_{30}\cos\vartheta_{30}\right)~.
\end{equation}
It should be noticed that for $\Xi\sim\Xi_{max}$, $k_3+k_3^\ast$ is a very
small real number. 
In the long wavelength regime the metric perturbation is essentially
driven by the expansion. 
For the computation of $\delta\varphi$ we use also eq.(\ref{a1}) and obtain
\begin{equation}
\delta\varphi=-\frac{k_2}{4\pi Ga}\dot{\phi}^\ast=-\frac{k_2}{4\pi
G}\phi^\ast_t~.
\end{equation}
For this case the source term gives only a negligible contribution to 
$\delta\varphi$, 
and $\psi$ gets another term proportional to $1/a$, which is absorbed in its
homogeneous solution. 

In the intermediate case, where $|\dot{h}-h^2|\ll k^2\ll
|\ddot{p}/p|$, 
$u$ is given by 
\begin{equation}
u=k_1\;p+k_2\;p\int \frac{d\eta}{p^2}-p\int
\frac{1}{p^2}\left[\int^{\eta}
\frac{k_{\psi}k^2}{2\dot{\phi}^{\ast}}
d\tilde{\eta}\right]\;d\eta~. \label{uinteros}	
\end{equation}
After some computations we see that the solution of $\psi$ is
also given by eq.(\ref{psilwos}). For the complex scalar field
perturbation instead we have
\begin{gather}
\delta\varphi=-\frac{k_2}{4\pi Ga}\dot{\phi}^\ast\mp \frac{ik_\psi
k^2}{36m^3 a}\dot{\phi}~.
\end{gather}
 
Hence, the first order perturbation of the scalar field 
decays in the long wavelength and in the intermediate regime, whereas
it oscillates around a constant value in the short wavelength. 
Therefore, during the oscillatory phase there can be no formation of
gravitational seeds, like boson stars, 
by a {\it massive} scalar field with 
sufficiently large bosonic charge.

Using the background asymptotic solution \cite{sci2:p2}, we 
obtain the analytic behavior of the metric and the complex
perturbations. They are reported in Table 1.
For all the other cases the solutions have to be found numerically. 
\section{Numerical results}\label{section5:p3}
To complete our analysis, we study the evolution of
the perturbations after inflation, which implies that the bosonic charge $\Xi$
is close to zero. 

As a first step we integrate the background equations from an initial
singularity point $S$, where the equation of state is given by
$\varepsilon^{(0)}=\mathbf{T^{(0)}}$ (stiff matter) and the asymptotic
behaviors of $\phi$, $\dot{\phi}$ and $H$ are as follows \cite{sci2:p2}
\begin{equation}
\phi=M_p\sqrt{3}\,
\left[C+\,i\,\frac{1}{3}\ln\left(t/t_0\right)\right]~,
\end{equation}
\begin{equation}
\dot{\phi}=M_p\sqrt{3}\,\left[-\frac{Cm^2t}{2}+\frac{i}{3t}\right]~,
\end{equation}
\begin{equation}
H=\frac{1}{3t}~,
\end{equation}
for $t\rightarrow 0^+$, $C$ and $t_0$ being integration constants. It
should be noticed that the bosonic charge $\Xi$ depends directly on
the choice of the value of  
$C$ and $t_0$. Hence, starting from $S$ we can freely fix the value of
$\Xi$. 
In order to have a long inflationary phase we must choose a small
value of $\Xi$, which is achieved for instance by setting 
$C=1$ and $t_0 =100$. With these values the background
starts with a stiff matter regime and, after an inflationary phase ($h$
exponentially increases, see Fig. 1),  ends in the
oscillatory phase ($h$ oscillates around a slowly decreasing curve, see
Fig. 1). For a massive complex scalar field inflation is
also less 
effective when $\Lambda$ increases (indeed, the maximum of $h$ decreases with
$\Lambda$, see Fig. 1). This, because the potential becomes
less flat and as a consequence the  
slow-roll approximation, needed for inflation to occur,
is no longer well satisfied. 

As a next step we integrate numerically eq.(\ref{deij}), 
eq.(\ref{dkg}) and its complex conjugate using the background
solution. We control our numerical computation by looking if
eqs.(\ref{de00})-(\ref{de0i}) propagate. Moreover, we check if we
recover the analytic solutions described in section \ref{section4:p3}.

The Bardeen potential $\psi$ behaves as predicted
during inflation and remains, up to a smooth oscillation, constant after
inflation. This behavior is valid for any value of the parameters $k$,
$m$ and $\lambda$. 

For a long period during inflation $|\delta\varphi|$ decreases and
thus, at first glance,  
does not behave as expected from our analytic analysis (see
Fig. 3). For the cases $\Lambda=0.1$ and  $m=0$ with
$\lambda=1$ we even get 
that during inflation $|\delta\varphi|$ is only decreasing. This may
seem to be in contradiction with the analytic results
obtained in 
section \ref{section4.1:p3}. However, this is not true since we do not
start the integration of 
the background right away from the singularity point, where inflation
occurs. If we did
so, we would have obtained the expected analytic
behavior for $|\delta\varphi|$. The $|\delta\varphi|$ trajectory which
we get by the analytic 
treatment, as given in Table 1, should be interpreted as an
``attractor'' for solutions. Indeed, if inflation is powerful (large e-fold
number), the time at which the perturbation is generated becomes
unimportant and $|\delta\varphi|$ will rapidly approach the behavior
found analytically. 
For the cases where inflation is not so powerful (see
Fig. 1), the shape of the fluctuations of  $|\delta\varphi|$
will depend more on the details of the background solution from which
they were 
produced. 
That is exactly what we find in
Fig. 3. After inflation the long wavelength perturbations 
of $|\delta\varphi|$ oscillate. 

For an initial fluctuation $\delta\varphi$, which lies inside the
Hubble radius, 
the exponential damping is so effective, 
that at the end of inflation $\delta\varphi$ practically vanishes, as can be
seen in Fig. 5. This is valid whatever 
the values of 
$\lambda$ and $m$ are. 

We notice that, to check numerically the analytic behavior for the
situation where $\Xi\sim\Xi_{max}$, is more difficult, since one has
to deal with numerical instabilities.

\section{Concluding Remarks}\label{section6:p3}

In this paper we studied the perturbation of the coupled
Einstein-Klein-Gordon equations for a complex scalar field in
different regimes. We derived analytically the time evolution of 
long and short wavelength perturbations during
inflation. Moreover, we established that in order to have a long
period of inflation the 
bosonic charge must be close to zero. After inflation the perturbation
of the metric remains constant during the oscillatory phase. 

If the complex scalar field has a large 
bosonic charge or the values of the parameters entering in the
potential are such that the slow-roll approximation does
not apply, then it turns out that inflation can not have taken place or
lasted only for a short period. 
Hence, the background passes directly in 
the oscillatory phase. We showed that in the massive case, for 
$\Xi\sim 0$ or $\Xi\sim\Xi_{max}$, the perturbations of the scalar
field decay or, at 
best, stay constant and thus no gravitational seeds, like boson stars, can 
form during this epoch. 
Constraints on the parameters of the scalar field 
potential coming from the measured CMB
anisotropies can only be put once a thorough study of the
quantized perturbations will be done. 
However, since 
inflation is essentially driven by one component of the 
complex field, we do not expect significant differences
to occur on the constraints and thus on the physical implications 
as compared to the real scalar field case \cite{kn:Mukhanov:p2}.

We also derived the Jeans wavenumber for the Newtonian regime
starting from the general relativistic equations. 

\section*{Acknowledgments}

We thank N.~Deruelle and N.~Straumann for very useful discussions.

\pagebreak

\section*{Caption}

\noindent Table 1: 

\noindent Summary of the analytic solutions during the inflationary and
the oscillatory phases for the metric and
the complex perturbations. 

\vskip 0.5cm
\noindent Figure 1: 

\noindent $h = a H $ as function of the cosmological time $t$
for three different values of $\Lambda = \frac{\lambda M_p^2}{m^2}$
and for the case $m=0$, $\lambda=1$.  
$\lambda$ is the coupling constant of the quartic 
self-interaction term, $m$ is the mass of the scalar field and $M_p$
is the Planck mass. For the massive cases $h$ ($t$) is in units
of $m$ ($m^{-1}$) and for the massless case $h$ ($t$)
is in units of $M_p$ ($M_p^{-1}$). 

\vskip 0.5cm
\noindent Figure 2:

\noindent Bardeen potential $\psi$ for the zero mode $(k = 0 )$ as a function 
of the cosmological time $t$
for different values of $m$ and $\lambda$. The units of $t$ are the
same as in 
Fig.~1.

\vskip 0.5cm
\noindent Figure 3:

\noindent Norm of $|\delta\varphi|$ for 
the zero mode as a function of the cosmological time $t$
for different values of $m$ and $\lambda$. For the massive cases 
(massless case) $|\delta\varphi|$ is in units of $m$
($M_p$). For the units of $t$ see Fig. 1. 

\vskip 0.5cm
\noindent Figure 4:

\noindent Same as for Fig. 2 but with $k = 60000$

\vskip 0.5cm
\noindent Figure 5:

\noindent Norm of $|\delta\varphi|$ 
for $k = 60000$ and $\Lambda =0$ as a function of the cosmological time
$t$. For the units of $|\delta\varphi|$ and $t$ see
Fig. 3 and Fig. 1.

\pagebreak

\vfill \indent
\begin{table}[!h]
\begin{center}
\begin{tabular}{|p{4.1cm}|p{3.5cm}|p{4.4cm}|} \hline
 \multicolumn{3}{|c|}{During Inflation}\\ \hline
 & $\lambda=0$  & $\lambda\neq 0$\\ \hline
$k^2\gg \ddot{g}/g$ 

(short wavelength) 
& 
$\psi$: oscillates 

$\delta\varphi$: \setbox3=\hbox{exponentially} \parbox[t]{\wd3}{
exponentially damped}
&
$\psi$: \setbox3=\hbox{oscillates in an
expo}\parbox[t]{\wd3}{oscillates in an exponentially decaying
\hfil\break envelope} 
 
$\delta\varphi$: \setbox3=\hbox{exponentially}
\parbox[t]{\wd3}{exponentially damped} \\ \hline
$k^2\ll \ddot{g}/g$ 

(long wavelength) 
& 
$\psi\propto\frac{1}{t^2}$ 

$\left|\delta\varphi\right|\propto -\frac{1}{t}$

($t\rightarrow -\infty$) 
&
\rule[-3mm]{0mm}{9mm} $\psi\propto\exp \left(4M_p\sqrt{\lambda /
3}\,t\right)$ 

$\left|\delta\varphi\right|\propto\exp
\left(2M_p\sqrt{\lambda / 3}\,t\right)$ 

($t\rightarrow -\infty$)\\ \hline 
 \multicolumn{3}{c}{}\\ \hline
 \multicolumn{3}{|c|}{During the Oscillatory Phase}\\ \hline
&\multicolumn{2}{|c|}{\rule[-3mm]{0mm}{7mm} $m\neq 0$ and
$\Xi/\Xi_{max}\approx 1$} 
\\ \hline
\vskip -2mm
\vbox{\hbox{$k^2\ll |\dot{h}-h^2|\ll
\ddot{p}/p$\hfil} \hbox{and\hfil} 
\hbox{$|\dot{h}-h^2|\ll k^2\ll\ddot{p}/p$\hfill}}

(long wavelength)
& \multicolumn{2}{|c|}{\parbox[t]{7.7cm}{\vskip -2mm
$\psi\propto 1/a \propto 1/t^{2/3}$ \vskip 0cm
$\left|\delta\varphi\right|\propto\frac{1}{t}$ \vskip 0cm
($t\rightarrow +\infty$)}} \\ \hline		
$k^2\gg \ddot{p}/p$ 

(short wavelength)
&\multicolumn{2}{|c|}{\parbox[t]{7.7cm}{\rule[-3mm]{0mm}{9mm}
$\psi$: \parbox[t]{6.8cm}{
2 terms, the first oscillates in a $\frac{1}{t}$ envelope, whereas the second
is proportional to $1/t^{2/3}$ for
$t\rightarrow +\infty$}
$\delta\varphi$: \parbox[t]{6.8cm}{2 terms, the first oscillates and the
second oscillates 
in a $\left(\frac{1}{t}\right)^{2/3}$
envelope, for $t\rightarrow +\infty$.\rule[-3mm]{0mm}{7mm}}}} \\ \hline
\end{tabular}
\vskip \baselineskip
\hbox{\hskip 7cm Table 1}\label{table1}
\end{center}
\end{table}
\vfill
\pagebreak

\vfill\indent
\begin{figure}[!h]
\begin{center}
\epsfig{file=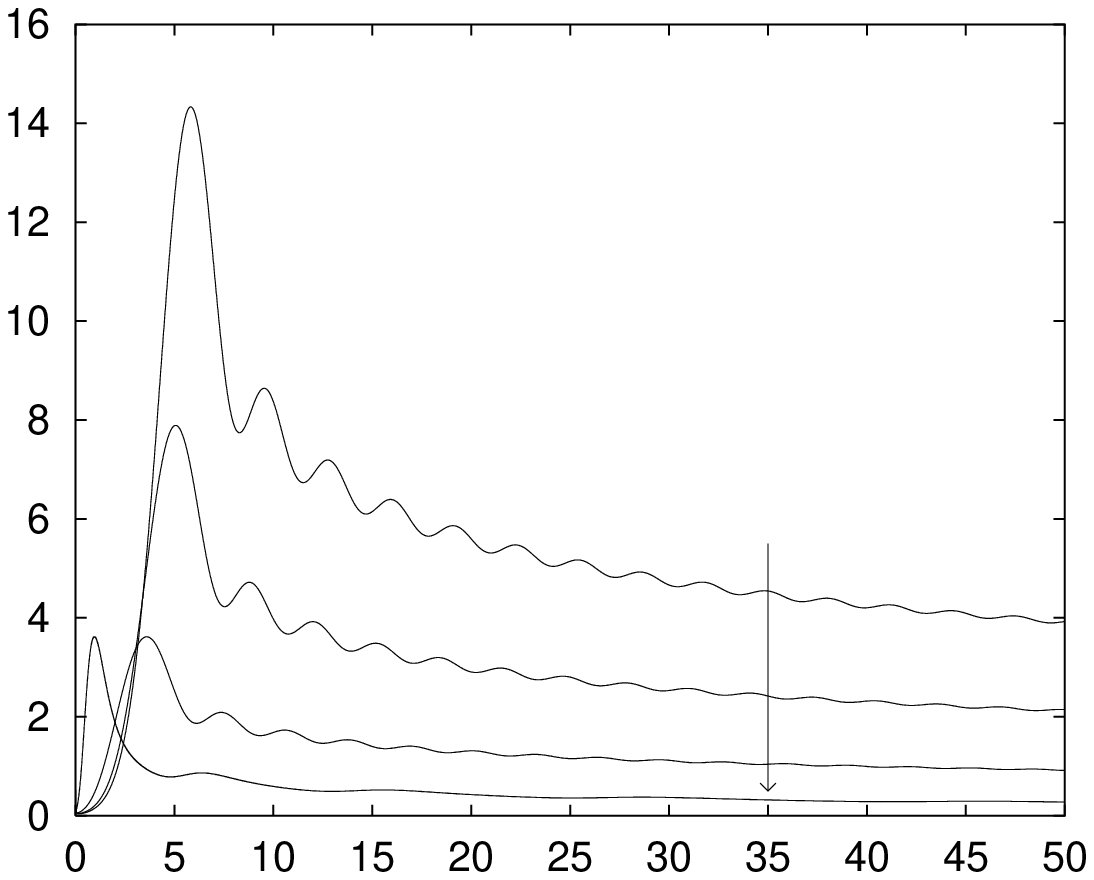,width=12cm}
\end{center}
\vskip -4cm
\hskip 1.5cm \begin{rotate}{-90}
             $10^{-4}\times h$  
             \end{rotate}
\vskip 4cm
\vskip -\baselineskip
\vskip -0.7cm
\hbox{\hskip 11.5cm $t$}
\vskip 0.7cm
\vskip -\baselineskip
\vskip -7cm
\hbox{\hskip 4.5cm $\Lambda = 0$}
\vskip 7cm
\vskip -\baselineskip
\vskip -4cm
\hbox{\hskip 4.2cm $\Lambda = 0.02$}
\vskip 4cm
\vskip -\baselineskip
\vskip -2.7cm
\hbox{\hskip 3.9cm $\Lambda = 0.1$}
\vskip 2.7cm
\vskip -\baselineskip
\vskip -4.3cm
\hbox{\hskip 8.5cm $\lambda = 1$, $m=0$}
\vskip 4.3cm
\vskip -\baselineskip
\hbox{\hskip 7cm Fig.1}\label{fig0:p3}
\end{figure}
\vfill
\pagebreak

\vfill\indent
\begin{figure}[!h] 
\begin{center}
\epsfig{file=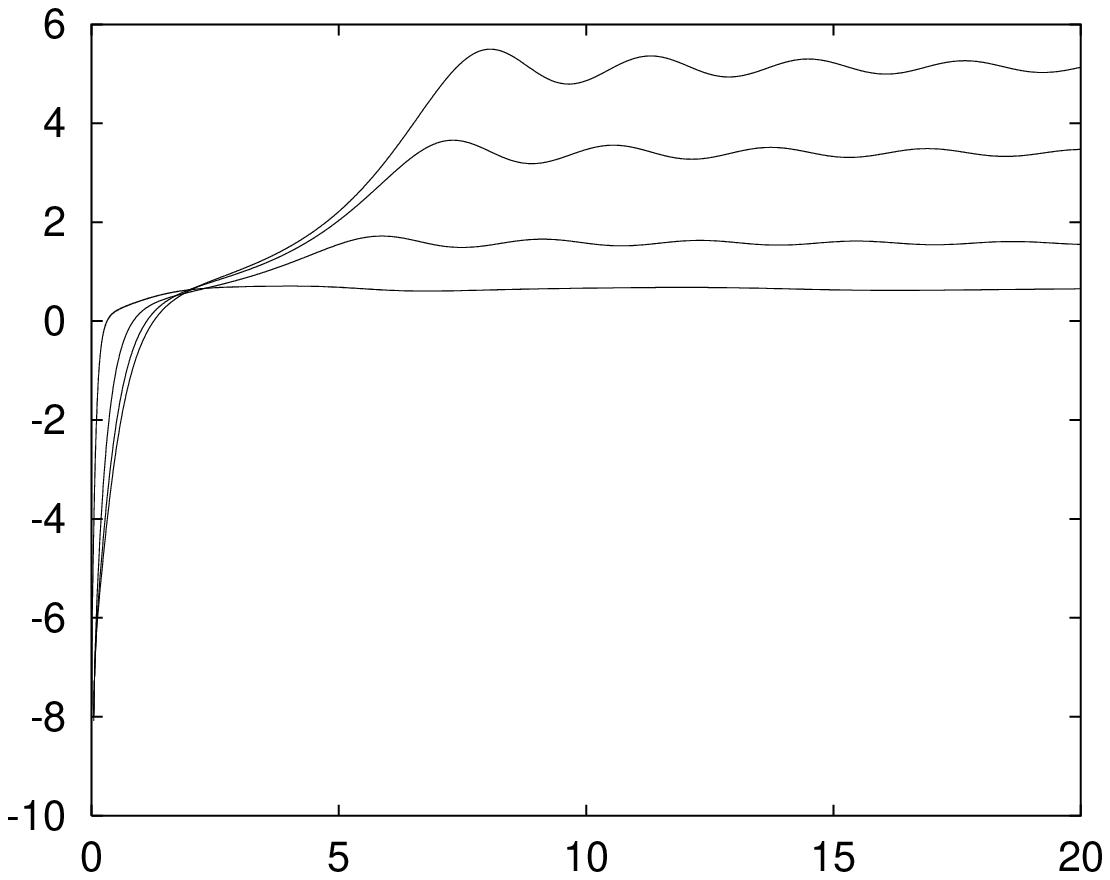,width=12cm}
\end{center}
\vskip -4.5cm
\hskip 1cm \begin{rotate}{-90}
$10^3\times\psi$
\end{rotate}
\vskip 4.5cm
\vskip -\baselineskip
\vskip -0.7cm
\hbox{\hskip 11.5cm $t$}
\vskip 0.7cm
\vskip -\baselineskip
\vskip -8.5cm
\hbox{\hskip 4.7cm $\Lambda = 0$}
\vskip 8.5cm
\vskip -\baselineskip
\vskip -7.5cm
\hbox{\hskip 7cm $\Lambda = 0.02$}
\vskip 7.5cm
\vskip -\baselineskip
\vskip -7.2cm
\hbox{\hskip 10cm $\Lambda = 0.1$}
\vskip 7.2cm
\vskip -\baselineskip
\vskip -6.3cm
\hbox{\hskip 9.5cm $\lambda = 1$, $m=0$}
\vskip 6.3cm
\vskip -\baselineskip
\hbox{\hskip 7cm Fig.2} \label{fig1pap3}
\end{figure}
\vfill
\pagebreak

\vfill\indent
\begin{figure}[!h]
\begin{center}
\vskip -1cm
\epsfig{file=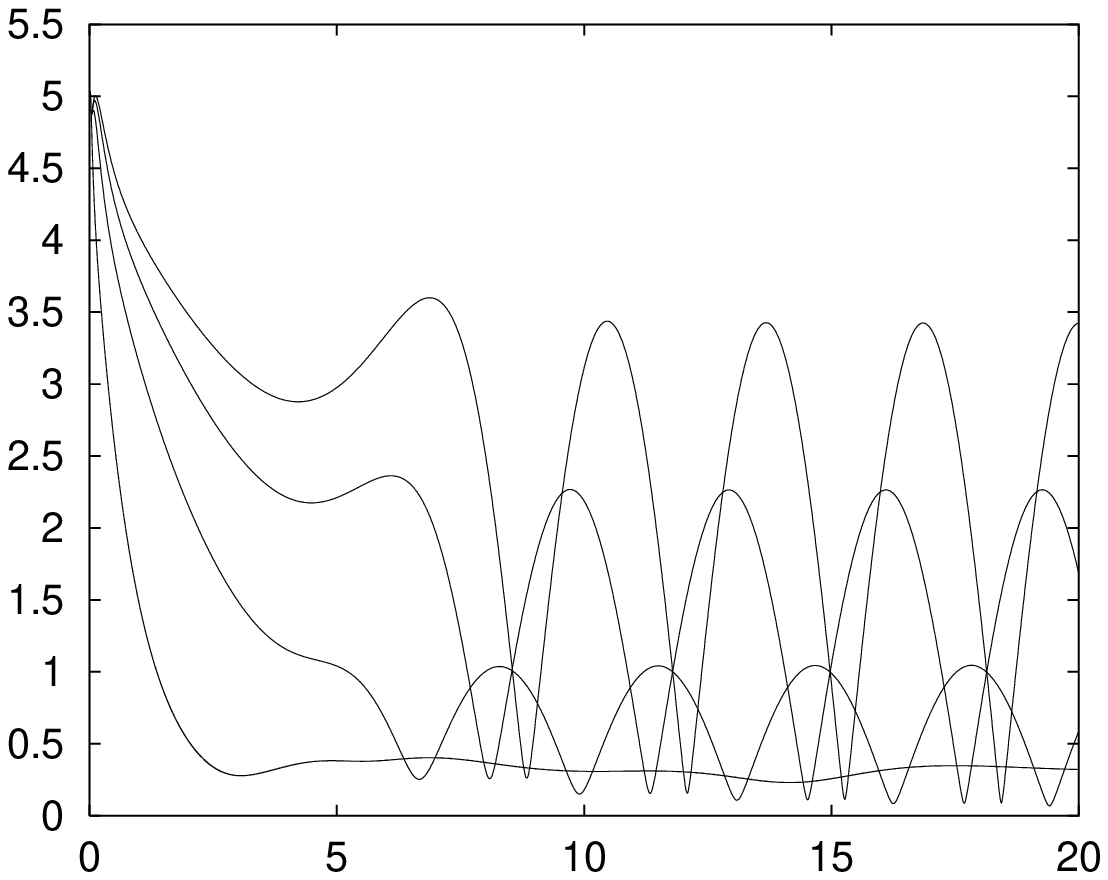,width=12cm}
\end{center}
\vskip -4.5cm
\hskip 1cm\begin{rotate}{-90}
          $10^3 \times |\delta\varphi|$
          \end{rotate}
\vskip 4.5cm
\vskip -\baselineskip
\vskip -0.7cm
\hbox{\hskip 11.5cm $t$}
\vskip 0.7cm
\vskip -\baselineskip
\vskip -6.8cm
\hbox{\hskip 5.7cm $\Lambda = 0$}
\vskip 6.8cm
\vskip -\baselineskip
\vskip -5cm
\hbox{\hskip 4.8cm $\Lambda = 0.02$}
\vskip 5cm
\vskip -\baselineskip
\vskip -3.5cm
\hbox{\hskip 4.4cm $\Lambda = 0.1$}
\vskip 3.5cm
\vskip -\baselineskip
\vskip -1.7cm
\hbox{\hskip 3cm $\lambda = 1$, $m=0$}
\vskip 1.7cm
\vskip -\baselineskip
\hbox{\hskip 7cm Fig.3}\label{fig2pap3} 
\end{figure}
\vfill
\pagebreak

\vfill\indent
\begin{figure}[!h]
\begin{center}
\epsfig{file=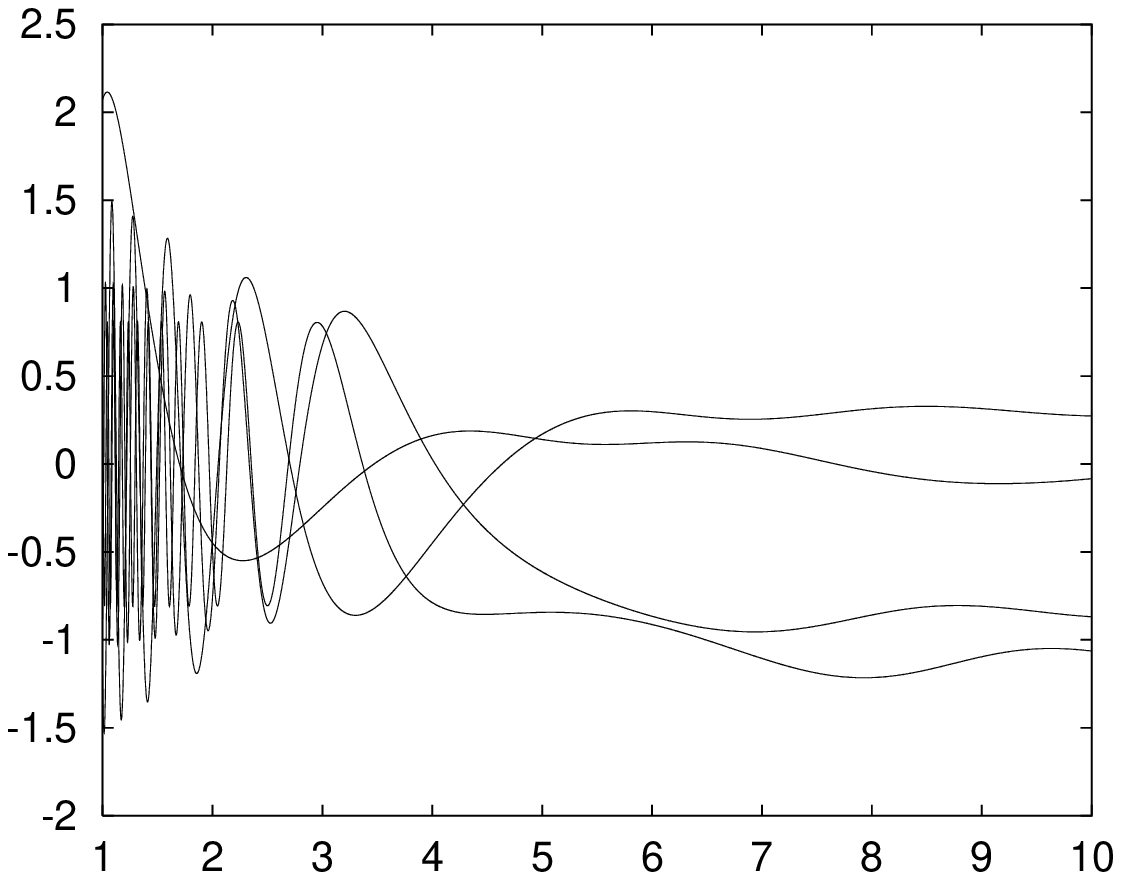,width=12cm}
\end{center}
\vskip -4.5cm
\hskip 1cm \begin{rotate}{-90}
	   $10^6\times\psi$
	   \end{rotate}
\vskip 4.5cm
\vskip -\baselineskip
\vskip -0.7cm
\hbox{\hskip 11.5cm $t$}
\vskip 0.7cm
\vskip -\baselineskip
\vskip -2.5cm
\hbox{\hskip 10cm $\Lambda = 0$}
\vskip 2.5cm
\vskip -\baselineskip
\vskip -3.7cm
\hbox{\hskip 8cm $\Lambda = 0.02$}
\vskip 3.7cm
\vskip -\baselineskip
\vskip -5.7cm
\hbox{\hskip 10.5cm $\Lambda = 0.1$}
\vskip 5.7cm
\vskip -\baselineskip
\vskip -4.4cm
\hbox{\hskip 9cm $m=0$, $\lambda = 1$}
\vskip 4.4cm
\vskip -\baselineskip
\hbox{\hskip 7cm Fig.4}\label{fig4pap3} 
\end{figure}
\vfill
\pagebreak

\vfill\indent
\begin{figure}[!h]
\begin{center}
\epsfig{file=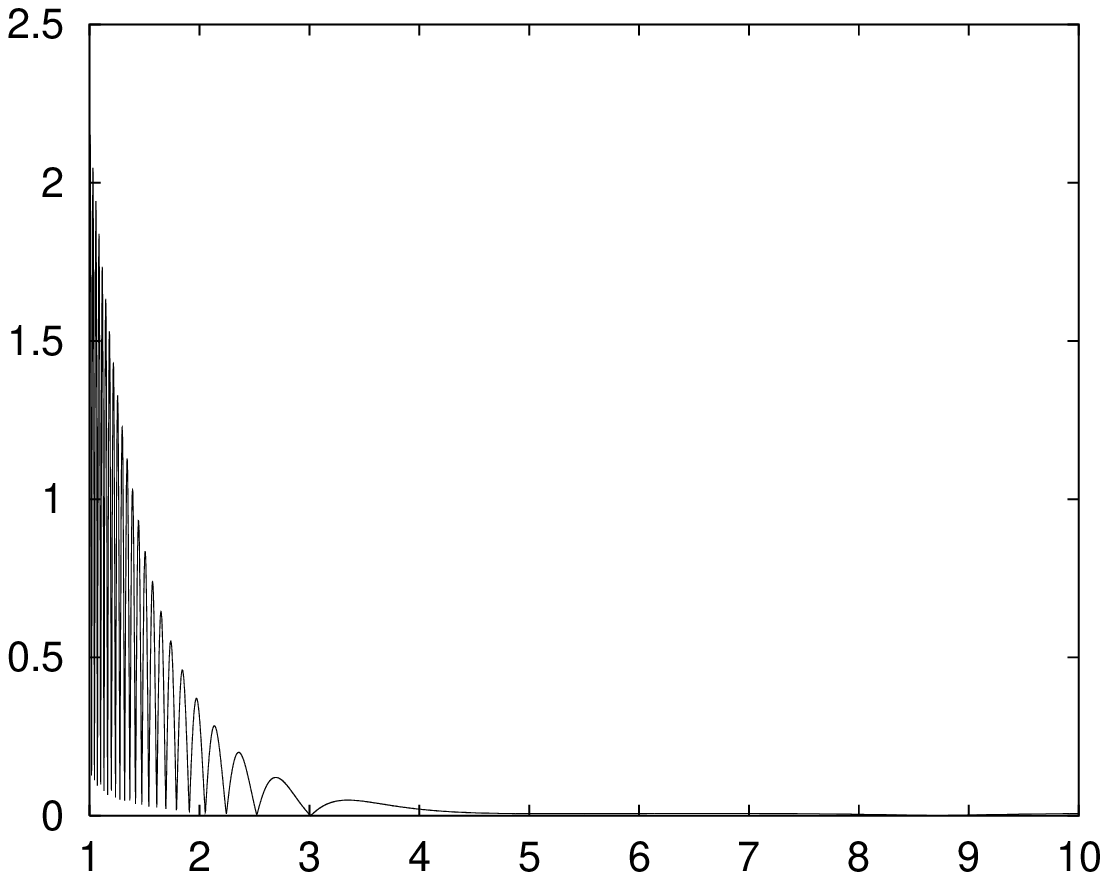,width=12cm}
\end{center}
\vskip -4.5cm
\hskip 1cm \begin{rotate}{-90}
	   $10^4\times |\delta\varphi|$
	   \end{rotate}
\vskip 4.5cm
\vskip -\baselineskip
\vskip -0.7cm
\hbox{\hskip 11.5cm $t$}
\vskip 0.7cm
\vskip -\baselineskip
\vskip -5cm
\hbox{\hskip 3.4cm $\Lambda = 0$}
\vskip 5cm
\vskip -\baselineskip
\hbox{\hskip 7cm  Fig.5 }\label{fig5pap3}
\end{figure}
\vfill
\end{document}